# Phenotypic class ratios as marker signs of different types of agamospermy

**Svetlana Sergeevna Kirikovich, Evgenii Vladimirovich Levites**

Institute of Cytology and Genetics, Siberian Branch of the Russian Academy of Sciences, Novosibirsk, Russia

Email: svetak@bionet.nsc.ru, elevites@ngs.ru

## ABSTRACT

This article focuses on the development of the method for the genetic classification of agamospermous reproduction types in plants using sugar beet as an example. The classification feasibility is ensured by the use of isozymes as genetic markers allowing the identification of all three phenotypic classes in the progeny of individual heterozygous diploid plant and is based on different phenotypic class ratios in the progenies obtained by meiotic and mitotic agamospermy. The data indicate that for sugar beet meiotic agamospermy is the more typical since 13 of 15 explored progenies were classified as those produced by meiotic agamospermy and only 2 as produced by mitotic agamospermy.

## 1. INTRODUCTION

The data obtained to date indicate that polymorphism in the agamospermous progenies of diploid plants is due to combinatorial processes. The combinatorial processes include meiosis occurring in tetraploid megaspore mother cells, which are present as admixture among the bulk of diploid cells of a mother plant [1].

Another combinatorial process occurs in apozygotic cells before its entering into embryogenesis. This process is a random equiprobable choice one allelic copy from each homologous chromosome among of many chromatid segments endoreduplicated as a result of polyteny [2-4]. The choice starts with random equiprobable attachment of one of the copies of each chromosome to the nuclear membrane or nuclear matrix. The unattached allelic copies are lost from the nucleus and the cell during the first stages of embryogenesis [2, 3]. This process is similar to the diminution in Cyclops [5, 6].

Joint action of these two combinatorial processes is also possible [4]. Diminution of excessive chromatin and subsequent entering into embryogenesis may be inherent to the cells of both the embryo sac and the nucellus or integuments [2-4, 7]. Different types of combinatorial processes lead to different phenotypic class ratios in agamospermous progenies [8].

Isozymes have become a convenient instrument for studying agamospermy. A wonderful peculiarity of isozymes is their codominant inheritance due to which the hybrid plant isozyme spectrum is different from each parent isozyme spectrum. For instance, one isozyme with fast (FF) or slow (SS) electrophoretic mobility, which corresponds to the genotype of a given locus, is revealed in the electrophoregram for the homozygote on the gene controlling this marker enzyme. But both enzyme allelic variants (isozymes) and also hybrid isozymes [9, 10] are revealed in the heterozygote (phenotype FS). This allows one to reveal all 3 phenotypic classes in the progeny of plant heterozygous at the isozyme locus: two homozygous (FF and SS) and one heterozygous (FS).

Using isozymes as genetic markers allows one to suggest the course of events leading to the appearance of such offspring, thereby to classify the types of agamospermous plant reproduction on the basis of phenotype ratios. Similar classification was proposed earlier. It takes into account the presence or absence of meiosis, thereby allowing the identification of meiotic and mitotic agamospermy by genetic methods [8].

With the appearance of evidence of the existence of combinatorial processes due to differential polyteny of chromosomes and diminution of redundant chromatid segments, it became possible to understand and classify the nature of agamospermy for a much larger number of offspring. In this connection, this work aims to give an interpretation of the previously obtained and new experimental data.

For this purpose, we studied phenotypic class ratios of marker enzymes in previously unexplored agamospermous progenies of sugar beet to evaluate the new possibility of using isozymes for the classification of agamospermy and to get a broader knowledge on the prevalence of agamospermy types in sugar beet.

## 2. MATERIAL AND METHODS

The agamospermous progenies obtained in 2003, 2004 and 2007 (Novosibirsk region) from pollensterile sugar beet plants in the pollenless regime were used in the investigation. Only plants of phenotypes ms0 and ms1 according to the Owen classification [11] were left in the field to create the pollenless regime during flowering. During flowering, the plants were repeatedly controlled for pollen sterility to discard the plants with a semisterile pollen phenotype (ms2) according to Owen's classification. Moreover, each experimental plant was covered with an unbleached calico isolator to prevent contamination by pollen.

The isozyme patterns of alcohol dehydrogenase (ADH1, EC 1.1.1.1), malic enzyme (ME1, EC 1.1.1.40), and isocitrate dehydrogenase (IDH3, EC 1.1.1.42) controlled by the loci *Adh1*, *Me1*, and *Idh3* respectively [12–14], were selected as marker traits. For electrophoretic analysis, each seed was freed from the pericarp and ground in 8-10 μl of the Tris–HCl buffer (pH 8.3) containing 292 mg of EDTA (disodium salt), 7.2 g of sucrose, and 0.3 g of dithiothreitol in 100 ml solution. Pieces of Whatman 3 MM chromatography paper (8x3mm) were soaked with the resulting homogenate and loaded into starch gel. The specimens were separated by horizontal electrophoresis in 14% starch gel with subsequent histochemical staining of electrophoregrams [14, 15]. The starch gel was prepared by boiling previously hydrolyzed commercial potato starch in buffer. For preparing gel buffer, the initial 0.75 M Tris–citrate buffer (pH 7.0) was dissolved 60-fold; for electrode buffer, 20-fold [16]. Electrophoresis was conducted at 110 V for 15 h at 4°C. On completion of electrophoresis, gel blocks were cut into 1-mm slices to stain the contained dehydrogenases by tetrazolium technique [17, 18]. For this purpose, the gels were incubated in the mixture containing 0.03 mg/ml phenazine metasulfate (PMS) and 0.15 mg/ml nitroblue tetrazolium in 0.05 M Tris–HCl buffer (pH 8.3). The reaction mixture for detecting ADH1 (15 ml) was supplemented with 1 mg of NAD and three drops of ethanol; the reaction mixtures for detecting ME1 and IDH3 (of the same volume) were supplemented with 1 mg of NADP, 0.2 ml of 1 M $MgCl_2$ solution, and the corresponding substrates (0.4 ml of 1 M sodium malate solution, 1–2 mg of sodium isocitrate, respectively). The gels were stained in a thermostat at 30°C for 1–3 h.

Comparison of marker enzymes phenotypic class ratios in the seeds obtained in different years was carried out using G-criterion [19]. The revealed experimental ratios were compared to the theoretically expected ones (3:8:3 and 1:2:1) using criterion $\chi^2$.

## 3. RESULTS AND DISCUSSION

Tables 1 and 2 show the results for agamospermous progenies obtained in different years. The phenotypic class ratios are compared with the ratio 3:8:3, which is typical of the gametic ratio in tetraploid plants. The comparison was made on the ground that the polymorphism detected in diploid agamospermous progeny obtained from a diploid plant is conditioned by meiosis in tetraploid cells, which are present as an admixture among the bulk of diploid cells of the maternal plant [1]. Also comparisons with the classical Mendelian ratio 1:2:1 have been made. One of all analyzed progenies (12-2) was obtained with the use of calico and parchment insulator. The phenotypic class ratios revealed by different methods of isolation were summated since they were not different when valued by G-criterion.

The phenotypic class ratio for ADH1 67FF: 94FS: 68SS in the progeny 1-4 differs significantly from 3:8:3 and 1:2:1 (**Table 1**). This ratio was published in our previous article [4] and was explained as follows. After the formation of egg cells in the ratio of 3:8:3, in the absence of pollen and pollination the polytenization of individual segments of chromosomes in egg cells may occur. But before entering into embryogenesis each egg loses excessive allelic copies of chromatid segments. This loss occurs randomly and equiprobably. As a result, in the nuclei of some initially heteroallelic egg cells only two identical alleles remain and the percentage of heteroallelic egg cells (*FS*) decreases and homoallelic egg cells (*FF* and *SS*) increases. The calculations of this process are described in detail in an earlier article [4]. It is noteworthy that in this case the heteroallelic class share decreases so strongly that its occurrence gets significantly lower than 0.5 and the ratio 1:2:1 gets disrupted (**Table 1**). In other cases, the reduction was less strong, and the phenotypic class ratios corresponded to 1:2:1 (**Table 1** and **Table 2**).

The data given in **Table 1** and **Table 2** indicate that a substantial portion of the presented progenies have symmetrical phenotype ratios or show insignificant deviations from the symmetry, which do not result in significant differences from 1:2:1. Based on our previous studies we can state that such progenies are obtained through the meiotic agamospermy pathway. This pathway includes the processes such as meiosis in tetraploid mother megaspore cells, polytenization in chromosomal regions carrying marker enzyme locus and random equiprobable diminution of redundant allelic copies in egg cells before its entering into embryogenesis. The symmetry in the phenotypic class ratio is due to the equal dose of alleles in heterozygous tetraploid megaspore mother cells maintaining after the duplication of the number of chromosomes in some diploid cells of a diploid maternal plant. For example, polyploidization of diploid cells having genotype *FS* leads to the appearance of tetraploid cells having genotype *FFSS* with an equal dose of alleles. This equality is preserved during subsequent polytenization of the regions of homologous chromosomes carrying this marker gene. It should be noted that, according to our estimates, the degree of polytenization of chromosome regions in egg cells is as low as 2 [4].

Another type of phenotypic class ratios is characterized by sharp differences from the ratio 1:2:1, which can be observed in the offsprings of plants 17-2 and 17-6 (**Table 2**). These ratios can also differ significantly from the ratio 3:8:3. The formation of progenies of plants 17-2 and 17-6 are more likely attributable to mitotic agamospermy as indicated by the pronounced asymmetry of their phenotypic class ratios (Table 2). This asymmetry may be due to a known independence of polytenization degree of different alleles in the somatic cells of plants, leading to a different dose of alleles [20].

<div style="text-align: right">**Table 1**</div>

Phenotypic class ratios for alcohol dehydrogenase (ADH1) and malic enzyme (ME1) in agamospermous progenies of sugar beet

| № of plant, year of reproduction | ADH1 | | ME1 | | Type of agamospermy |
|---|---|---|---|---|---|
| | FF:FS:SS | 1) $\chi^2$(3:8:3) 2) $\chi^2$(1:2:1) | FF:FS:SS | 1) $\chi^2$(3:8:3) 2) $\chi^2$(1:2:1) | |
| 1-1 (2003) | 33:52:9 | 1)**14.4255\*\*\*** 2)**13.3.3191\*\*** | 25:32:17 | 1)**7.856\*** 2)3.081 | Meiotic |
| 11-1 (2003) | 73:0:0 | | 24:47:22 | 1)1.757 2)0.097 | Meiotic |
| 11-1 (2004) | 59:0:0 | | 19:55:26 | 1)1.331 2)1.980 | Meiotic |
| 2-6 (2003) | 14:33:16 | 1)0.731 2)0.270 | 25:0:0 | | Meiotic |
| 2-6 (2004) | 61:134:62 | 1)2,636 2)0,48 | 172:0:0 | | Meiotic |
| 2-8 (2004) | 26:29:15 | 1)**11.092\*\*** 2)5.514 | 20:29:21 | 1)**7.092\*** 2)2.086 | Meiotic |
| 12-2 (2003) | 30:46:22 | 1)5.690 2)1.673 | 27:49:22 | 1)2.637 2)0.510 | Meiotic |
| 12-2 (2004) | 23:32:23 | 1)**8.274\*** 2)2.512 | 25:31:22 | 1)**9.911\*\*** 2)3.513 | Meiotic |
| 7-12 (2003) | 35:0:0 | | 34:55:28 | 1)5.625 2)1.034 | Meiotic |
| 5-7 (2007) | 41:83:41 | 1)3,152 2)0,006 | 50:75:45 | 1)**12,120\*\*** 2)2,447 | Meiotic |
| 1-4 (2007) | 67:94:68 | 1)**24,233\*\*\*** 2)**7,349\*** | | | Meiotic |

Probability of affinity with theoretically expected ratios: *- $P<0.05$; **- $P<0.01$; ***- $P<0.001$

The asymmetric phenotypic class ratios under mitotic agamospermy can result also from the heterogeneity of the nucellus and integuments cells due to mixoploidy. Mixoploidy, or the presence of cells of different ploidy levels within the same tissue, has been demonstrated in many plant species, including sugar beet [21].

In our earlier work, the formation of progenies with only two detectable phenotypic classes with active enzyme was attributed to the mitotic agamospermy [22]. The presence of only two phenotypic classes is an extreme case of asymmetry. At the present time the existence of differential polyteny is an argument to attribute the progeny with three detectable phenotype classes, with normal enzymatic activity and asymmetric phenotype class ratios to the mitotic agamospermy. The formation of progenies with a high heteroallelic class share, which is significantly greater than for the ratio 3:8:3, could also be attributed to mitotic agamospermy.

Phenotypic class ratios for isocitrate dehydrogenase (IDH3) and malic enzyme (ME1) in agamospermous progenies of sugar beet

| № of plant, year of reproduction | IDH3 | | ME1 | | Type of agamospermy |
|---|---|---|---|---|---|
| | FF:FS:SS | 1) $\chi^2(3:8:3)$ 2) $\chi^2(1:2:1)$ | FF:FS:SS | 1) $\chi^2(3:8:3)$ 2) $\chi^2(1:2:1)$ | |
| 11-2 (2004) | 31:69:47 | 1)**10,313**\*\* 2)4,034 | 45:60:44 | **1)17,34**\*\*\* 2)5,658 | Meiotic |
| 12-5 (2003, 2004) | 21:26:20 | 1)**9,2338**\*\* 2)3,3881 | 18:40:15 | 1)0,452 2)0,918 | Meiotic |
| 17-2 (2003) | 34:30:11 | 1)**25.458**\*\*\* 2)**17.107**\*\*\* | 40:0:0 | | Mitotic |
| 17-6 (2003, 2004) | 52:51:17 | 1)**34.326**\*\*\* 2)**23.117**\*\*\* | 36:0:0 | | Mitotic |

Probability of affinity with theoretically expected ratios: \*- $P<0.05$; \*\*- $P<0.01$; \*\*\*- $P<0.001$

Diminution of redundant allele copies can be considered as a specific mechanism in seed reproduction. In somatic cells, a surplus of genetic material arising, for example, through polyploidy is compensated in some extent by gene inactivation [23]. Similar inactivation can occur also at the polytenization of chromosomes in somatic cells. It was previously hypothesized that polytenization might initiate a number of transformations of the genome, including the movement of transposable elements and chromosome DNA methylation [3]. The question arises whether there is any connection between the diminution of redundant copies of genes and the methylation of their alleles and whether methylation can have an impact on the phenotypic class ratios in agamospermous progenies? To elucidate this topical question is an interesting task and can help in considering all the factors influencing polymorphism in agamospermous progenies.

The suggested genetic method for the classification of agamospermous reproduction types differs from the cytological approach by providing an opportunity to analyze a large number of plants, thereby claiming to be more objective. On the other hand, the genetic method does claim to complement rather than supersede the cytoembryological methods.

The data indicate that for sugar beet meiotic agamospermy is the more typical since 13 of 15 explored progenies were classified as those produced by meiotic agamospermy and only 2 as produced by mitotic agamospermy.